\documentclass{ws-procs9x6}
\newcommand{\leplep}{\ell^{+}\ell^{-}}
\newcommand{\jp}{J/\psi}
\newcommand{\jpsi}{J/\psi}

\newcommand{\pipi}{\pi^{+}\pi^{-}}

\newcommand{\rt}{\rightarrow}
\newcommand{\etal}{\em et al.}

\begin{document}
\title{Properties of the X(3872) }
\author{S.-K.Choi \\
for the Belle collaboration }
\address{Department of Physics, College of Natural Sciences, \\
900 Gazwa-dong, Jinju, 660-701, Republic of Korea\\ 
E-mail: schoi@gsnu.ac.kr}
\maketitle
\abstracts{
We report recent results on the properties of the X(3872) produced
via the $B^{+} \rt K^{+} X(3872)$ decay process in the Belle detector. 
We present constraints on possible 
charmonium-state assignments for this particles.}

\section{Introduction}
A first step in understanding the X(3872) particle that was recently
discovered by Belle\cite{s-s} is to determine its $J^{PC}$ 
quantum numbers.  Here, we survey possible assignments and properties 
of the most likely candidates.
 We restrict our considerations
to $0^{++}$ and $1^{--}$ $\pipi$ systems\cite{close} and scenarios 
where the relative 
orbital angular momentum of the $\pipi$ and $\jp$ is $L \le 3$.
%
In this report we concentrate on possible charmonium assignments,
and only those where decays to 
$D\overline{D}$ are forbidden or expected to be strongly suppressed.
For the case of a $0^{++}$ dipion, there are three charmonium 
possibilities:
the $h_c ^{'} (2^1P_1)$
and two triplet D-wave states, the $\psi_2 (^3D_2)$ and 
$\psi_3 (^3D_3)$.
For the $1^{--}$ dipion case, there are also three possibilities: the 
$\eta_{c} ^{''}$, the $\chi_{c1} ^{'}$ and the $\eta_{c2} (^1D_2)$.
For these assignments, the $\pipi \jp$ decay would violate isospin and 
should be strongly suppressed.
\section{Search for X(3872)$\rt \gamma \chi_{c2}$ ($\chi_{c1}$)}
The Wigner-Eckart theorem says that 
the widths for $\psi_{2} \rt \pipi \jp$ and $\psi_3 \rt \pipi \jp$ should 
both be equal to $\Gamma (\psi (3770) \rt \pipi \jp)$.
This has been recently measured by BESII\cite{bes2} and CLEO-c\cite{cleoc}
to be $80 \pm 32 \pm 21$ keV and $\le 55$ keV (90$\%$ CL), respectively.
The results are in some contradiction with each other. For the following
discussion we conservatively assume an upper limit derived from
the larger BES number of $\Gamma (\psi (3770) \rt \pipi \jp) < $ 129 keV.
   
Calculations of the $\gamma \chi_{c1}$ width for an M=3872 MeV $\psi_2 $
range from 207~keV\cite{eichten} to
360~keV\cite{barnes}.
The 90\% CL upper limit of 
\begin{eqnarray}
 \frac {\Gamma (X \rt \gamma \chi_{c1})}{\Gamma (X \rt \pipi \jp)} < 0.89
\end{eqnarray}
that was reported in Ref.[1] contradicts these expectations for the 
$\psi_2$.

Barnes and Godfrey\cite{barnes} observe that 
although $\psi_{3} \rt D \overline{D}$
is allowed for a 3872 MeV $\psi_{3}$, this mode is suppressed by an $L=3$ 
centrifugal 
barrier and the total $\psi_{3}$ width may be less than the $\Gamma<2.3$ 
MeV experimental upper limit\cite{s-s}.
These authors, and also Eichten, Lane and Quigg\cite{eichten}, propose 
the $\psi_{3}$ as a charmonium candidate for the X(3872).

%
For an M=3872~MeV $\psi_3$, the calculated $\gamma \chi_{c2}$ widths 
range from 299~keV\cite{eichten} to 370~keV\cite{barnes}. 
Thus, the partial width for 
$\psi_{3} \rt \gamma \chi_{c2}$ is expected to be more than
twice that for $\psi_{3} \rt \pipi \jp$.
We performed a search for $X \rt \gamma \chi_{c2}$ that followed 
closely the procedure used for the $\gamma \chi_{c1}$ limit 
reported in Ref.[1].  We require one of the $\gamma \jp$ combinations to 
satisfy 444~MeV $<(M_{\gamma \leplep} - M_{\leplep} ) <$ 469~MeV. 
The $M_{bc}$ and $\Delta E$ signal regions are $|M_{bc} - 5.28|<$ 0.009 GeV
and -0.04 $< \Delta E <$ 0.03 GeV.

\begin{figure}[ht]
\centerline{\epsfxsize=4.1in\epsfbox{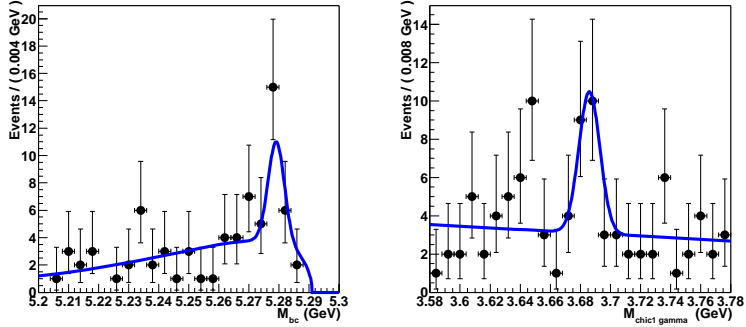}}   
\caption{Signal-band projections of $M_{bc}$ (left) and $M_{\gamma \chi_{c2}}$ 
(right) for events in the $\psi^{'}$ region with the 
results of the unbinned fit superimposed.
\label{psip2gammachic2}}
\end{figure}

We use the $B \rt K \psi^{'}; \psi^{'} \rt \gamma \chi_{c2}$ decay chain as a 
normalization reaction. The signal-band projections of $M_{bc}$ and 
$M_{\gamma \chi_{c2}}$ for the $\psi^{'}$ region are shown 
in Figs.~\ref{psip2gammachic2} (left) and (right), respectively, together 
with curves that show the results of the fit. The fitted signal yield is
$18.3 \pm 5.2$ events, where, based on known branching fractions,
we expect $12\pm 3$ events.

\begin{figure}[ht]
\centerline{\epsfxsize=4.1in\epsfbox{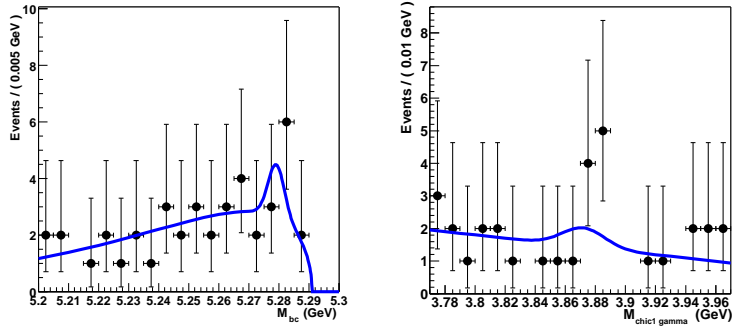}}   
\caption{Signal-band projections of $M_{bc}$ (left) and $M_{\gamma \chi_{c2}}$ 
(right) for events in the X(3872) region with the 
results of the unbinned fit superimposed.
\label{x2gammachic2}}
\end{figure}

Figure~\ref{x2gammachic2} show the same projections for events 
in the X(3872) region, where
there is no apparent signal. An unbinned fit produces a signal yield of 
$2.9 \pm 3.0 \pm 1.5$ events, where the first error is statistical and the 
second systematic. The latter is estimated by the changes that occur when 
the input parameters to the fit are varied over their allowed range of values.

The ratio of the $X \rt \gamma \chi_{c2}$ and the $X \rt \pipi \jp$ partial 
widths and its 90$\%$ CL upper
limit are
\begin{eqnarray}
 \frac {\Gamma (X \rt \gamma \chi_{c2})}{\Gamma (X \rt \pipi \jp)} 
= 0.42 \pm 0.45 \pm 0.23 < 1.1 ( 90\% CL),
\end{eqnarray}
where the second quoted error is the quadratic sum of the systematic 
uncertainties in acceptance, the branching fractions and variations in the 
$\gamma \chi_{c2}$ event yield for different fitting methods.


\section{Search for $X \rt \gamma \jp$}
The $\chi_{c1}^{'}$ is expected to be near 3968 MeV, well above the 
$D \overline{D}^{*}$ threshold, and its width is expected to be hundreds
of MeV\cite{eichten}.  If potential models are wrong and the
$\chi_{c1}^{'}$ is below the $D \overline{D}^{*}$ threshold
at 3872~MeV, it could conceivably be narrow and $\pipi \jp$ decays 
might be significant, even though these would violate isospin.
In this case, the $\gamma \psi^{'}$ and $\gamma \jp$ transitions would
be important and almost certainly have larger partial widths than that 
for the $\pipi \jp$ mode. We searched for the $X \rt \gamma \jp$ decay mode.

\begin{figure}[ht]
\centerline{\epsfxsize=4.8in\epsfbox{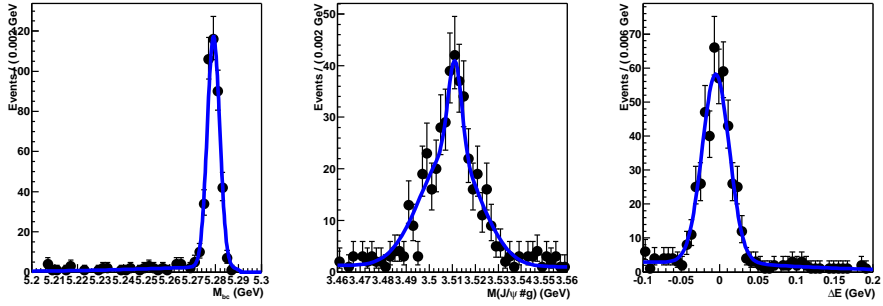}}   
\caption{Signal-band projections of $M_{bc}$ (left), $M_{\gamma \jp}$ (center)
and $\Delta E$ (right) for events in the $\chi_{c1}$ region with the 
results of the unbinned fit superimposed.
\label{chi12gammajpsi}}
\end{figure}
\begin{figure}[ht]
\centerline{\epsfxsize=4.8in\epsfbox{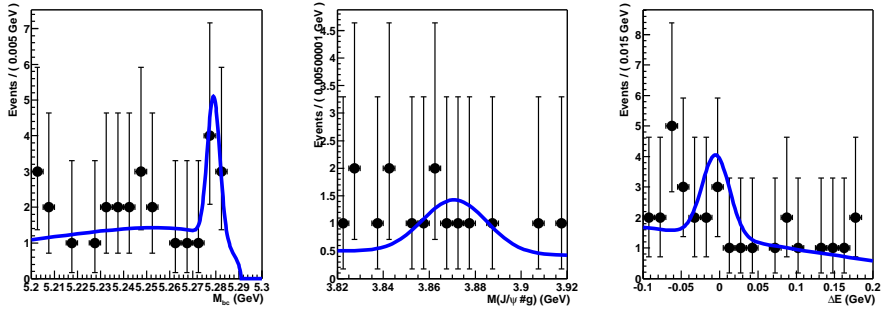}}   
\caption{Signal-band projections of $M_{bc}$ (left), $M_{\gamma \jp}$ (center)
and $\Delta E$ (right) for events in the X(3872) signal region with the 
results of the unbinned fit superimposed.
\label{x2gammajpsi}}
\end{figure}

We select $B^{+} \rt K^{+} \gamma \jp$ event candidates using 
the criteria given in ref.[1]. 
The $B^{+} \rt K^{+} \gamma \jp$ channel is dominated by 
$B^{+} \rt K^{+} \chi_{c1}$; $\chi_{c1} \rt \gamma \jp$ decays and we use
this as a calibration reaction. 
We define a $\chi_{c1}$ window for $\gamma \jp$ masses within 20 MeV of the 
nominal $\chi_{c1}$ mass. Figure~\ref{chi12gammajpsi} shows the signal-band 
projections for $M_{bc}$ (left), 
$M_{\gamma \jp}$ (center) and $\Delta E$ (right) 
for events in the $\chi_{c1}$ region with the results of a 
three-dimensional unbinned fit superimposed. 
The fitted number of events is $470\pm 24$.

We define an $X \rt \gamma \jp$ signal region to be 
$|M(\gamma \jp)-3872~{\rm MeV}|
<26$ MeV. Figure~\ref{x2gammajpsi} shows the same 
projections for events in the X(3872) signal region. 
Here there is no strong evidence for a signal: the fit gives a
$2.2\sigma$ signal yield  of $7.7\pm 3.6$ events.
The resulting limit is 
\begin{eqnarray}
 \frac {\Gamma (X \rt \gamma \jp)}{\Gamma (X \rt \pipi \jp)} 
= 0.22 \pm 0.12 \pm 0.06 <0.40 ( 90\% CL),
\end{eqnarray}
where the second quoted error is systematic and includes uncertainties in 
acceptance, the branching fractions and variations in the $\gamma \jp$ 
event yield for different fitting methods.

\section{Helicity angle distribution for $1^{+-} h_{c}^{'}$}
\begin{figure}[ht]
\centerline{\epsfxsize=3.1in\epsfbox{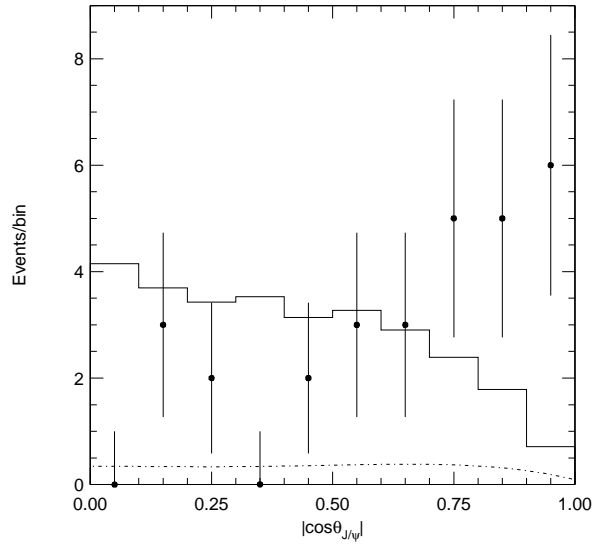}}   
\caption{The measured $| \cos\theta_{\jpsi} |$ distribution. 
The superimposed histogram is the normalized MC distribution
for the 1+- hypothesis. Here $\chi ^{2} / dof$ = 75/9.
\label{helicity1+-}}
\end{figure}

We define $\theta_{\jp}$  as the angle between the $\jp$ and 
the negative of the $K^{+}$ momentum vectors in the X(3872) rest frame
in the decay $B \rt X K; X \rt \pipi \jp$. 
The $|\cos\theta_{\jp}|$ distribution for
X(3872) events with $m_{\pipi}>$ 0.65 GeV is shown as
data points in Fig.~\ref{helicity1+-}. The smooth dotted curves are
polynomials that are fit to sideband-determined backgrounds.

Figure~\ref{helicity1+-} shows a comparison of the measured 
$|cos \theta_{\jp}|$ distribution 
with a MC sample generated with a $J^{PC}=1^{+-}$ hypothesis. 
Here the expected $|\cos \theta_{\jp}|$ distribution has 
a $\sin^2 \theta_{\jpsi}$ dependence that goes to zero at
$\cos \theta_{\jp} = 1$, where the data tend to peak. 
This makes the overall $\chi^{2}$ quite poor, $\chi^{2} / dof$ is 75/9, 
and enables us to rule out the $1^{+-}(h_{c} ^{'})$
hypothesis for the X(3872) with high confidence.

\section{Summary}
None of the six charmonium candidate states comfortably fit the
measured properties of the X(3872).
The $90\%$ CL branching fraction upper limit for $\it B (X(3872)
\rt \gamma \chi_{c2}$) decay is 1.1 times that for $\pipi \jp$. 
This conflicts with theoretical expectations for the case where the X(3872)
is the $3^{--} \psi_{3}$.

The possibility that the X(3872) is the $1^{++} \chi_{c1}^{'}$ is made
improbable by the limit 
$\it B (X \rt \gamma \jp) < 0.4\it B (X \rt \pipi \jp)$.
The former would be an allowed E1 transition 
with an expected width of $\Gamma_{\gamma \jp} \sim 10$~keV\cite{barnes}.
The latter would be an isospin-violating transition; other isospin 
violating 
transitions in the charmonium system have widths that are less than 1 keV.

An analysis of the $\theta_{\jp}$ helicity angle distribution 
eliminates the $1^{+-}(h_{c}^{'})$ hypothesis with a high degree of
confidence.

The $0^{-+}(\eta_{c}^{''})$ mass differs from that of the 
$\psi(3S)$ by hyperfine splitting and can be reliably expected to be about 
50 MeV (or less) below that of the $\psi(3S)$, which is at 4030~MeV.
Moreover, even if it were as low as 3872 MeV, the width is expected to be
some 10's of MeV, similar to that of the $\eta_{c}$ and wider than the
2.3~MeV upper limit for the X(3872). 
For $2^{-+}(\eta_{c2})$, the $\eta_{c2} \rt \pipi \eta_{c}$ and $\gamma h_{c}$
decays are allowed and expected to have widths in the range of 100's of 
keV\cite{barnes}, and much larger than that for the isospin-violating 
$\pipi \jp$ mode.  If the X(3872) were the $\eta_{c2}$, 
the total exclusive
branching fraction for the $B^{+} \rt K^{+} \eta_{c2}$ decay, 
which is non-factorizable and suppressed by an $L=2$ barrier, 
would be anomalously large.

\end{document}